\def\bibitemn#1#2{\bibitem{#1}#2}
\documentclass[epsfig,twocolumn,pra,aps,floats]{revtex4}

\def\comment#1{}

\def\p{{\rm p}}

\newcommand{\BF}[1]{\mbox{\boldmath $#1$}}

\begin{document}
\title{
World Nematic
Crystal Model of Gravity \\ Explaining the Absence of Torsion
 }
\author{H.~Kleinert}
 \affiliation{  Institut f\"ur Theoretische Physik.
 Freie Universit\"at Berlin, Arnimallee 14, D-14195 Berlin}
\author{J.~Zaanen}

            \affiliation{Instituut-Lorentz for Theoretical Physics,
  Leiden University, P.O.Box 9504, 2300 RA Leiden, The Netherlands }
%
%
\begin{abstract}
%
%
%
Assuming that at small distances space-time is
equivalent to an elastic medium which is isotropic in space and
time directions, we demonstrate that the quantum nematic liquid
arising  from this crystal by  spontaneous proliferation of
dislocations corresponds with a medium which is merely carrying
curvature rigidity. This medium is at large distances indistinguishable
from Einstein's spacetime of general
relativity. It does not support
torsion and possesses string-like curvature sources which
in spacetime form
world surfaces.
\end{abstract}

\maketitle

The 19-th century idea that space-time is filled with an elastic
medium called the ether seemed once and for all defeated by
 Einstein's relativity principle. However, it revived
in a  modern setting by the
realization that the theory of crystals containing topological defects
can be reformulated in the language of differential
geometry\cite{Kondo}.
In this language, the physics of the {\em plastic} medium has
striking formal resemblances with the theory of general
relativity. Dislocations and disclinations represent
torsion and curvature in the geometry of spacetime, as
discussed in  Refs.~\cite{paper172,NHOL,NHOL2,padua}
and the textbook \cite{GFCM}.
The plastic medium
provides us
with an elementary example of the idea of {\em emergent}
Lorentz invariance, advanced recently
 by \cite{Volovik1,Chapline,Zhang}.
\comment{We shall see that so far, there is no experimental
evidence against
 Euclidean space-time being
a nematic crystal of Planckian lattice spacing,
whose dynamics at large distances
is follows
a Lorentz invariant theory.}

The crystal-spacetime analogy is, however, not perfect. First,
in ordinary first gradient elasticity,
curvature sources are confined, since the forces between disclinations
grow linearly with the distance.
 One of us [HK] suggested some time ago \cite{paper172}
to cure  this
problem by assuming the world
crystal to be somewhat exotic, possessing
only second-gradient elasticity, but this
obscures the analogy. A  more fundamental problem
in such an elastic medium is that
torsion
is deconfined, while
space-time must be free of torsion
to be compatible with elementary particle physics
\cite{271.}. Remarkably,
these two problems find a natural solution by cross fertilization
with recent developments in condensed matter physics. Kivelson
et al. \cite{Kivelson}
suggested some time ago that quantum liquids exhibiting
liquid crystalline orders might exist, which are of potential
relevance in both quantum Hall physics \cite{quantumhall}
and high-$T_c$ superconductivity \cite{Zaanen1}.
This stimulated one of us [JZ] to construct the dynamical generalization
in 2+1 dimensions of the dualities underlying the theory
of classical melting in two dimensions \cite{HN,GFCM}.
In this non-relativistic context,
quantum nematic
states appear as Higgs condensates of dislocations which are at the
same time conventional superfluids \cite{Zaanen}.
In this note we shall demonstrate that
 the relativistic extension of such a
quantum nematic state describes a medium carrying merely curvature
rigidity, becoming  indistinguishable from Einstein's spacetime at
distances large compared to
 lattice spacing which is assumed to be
 Planckian.

Our model will be formulated as before \cite{paper172} in three euclidean
dimensions, for simplicity. The generalization to four dimensions
is  straightforward.
The elastic energy
is
expressed in terms of a material {\em displacement field\/}  $u_i({\bf x})$
as
\begin{equation}
 E =\int d^3x\, \left[ \mu
u_{ij}^2 ({\bf x})
+\frac{ \lambda }{2}u_{ii}^2  ({\bf x}) \right] ,
\label{@orig}\end{equation}
where
\begin{equation}
  u_{ij}({\bf x}) \equiv \frac{1}{2} [\partial _i u_j({\bf x}) +  \partial _j u_i({\bf x})]
\label{@}\end{equation}
 is the {\em strain tensor\/}
and
 $\mu, \lambda $ are the shear modulus and the Lam\'e constant, respectively.
 The elastic
 energy goes to zero for infinite wave length since in this
 limit $ u_i({\bf x})$ reduces to a pure translation under which
the
 energy of the system is invariant. The
crystallization process
causes a spontaneous breakdown of the
translational symmetry of the system.
The elastic distortions describe the Nambu-Goldstone modes
resulting from this symmetry
breakdown.
Note that so far the crystal has an extra
longitudinal sound wave with a different  velocity than the shear waves.

A crystalline material always contains defects. In their presence,
the elastic energy is
\begin{equation}
  E = \int d^3 x \left[
\mu (u_{ij} - u^{\rm p}_{ij})^2+
\frac{ \lambda }{2} (u_{ii} - u^{\rm p}_{ii})^2\right],
\label{@elen}\end{equation}
 where $u_{ij}^\p$ is the so-called {\em plastic strain
tensor\/}  describing the defects. It is composed of an ensemble
of
 lines with a {\em dislocation} density
\begin{equation}
    \alpha _{il} = \epsilon _{ijk }
      \partial _j \partial _k u_l({\bf x}) =  \delta _i
     ({\bf x};L) (b_l +  \epsilon _{lqr}  \Omega _q x_r).
\label{DLD}\end{equation}
and a {\em disclination} density
\begin{equation}
 \theta _{il} =  \epsilon _{ijk} \partial _j \phi^{\rm p}_{kl}
   =  \delta _i ({\bf x};L)  \Omega _l,
\label{DED}\end{equation}
where $b_l$ and $\Omega_l$ are the so-called Burgers and Franck
vectors of the defects. The densities satisfy the conservation
laws
\begin{equation}
\partial _i  \alpha _{ik} = -  \epsilon _{kmn}
    \theta _{mn},
    ~~~
      \partial _i   \theta _{il} = 0.
\label{DED}\end{equation}
 Dislocation lines are either closed or they end in disclination
 lines, and
 disclination lines are closed.
These are Bianchi identities
 of the defect
 system.

An important geometric quantity characterizing dislocation
 and disclination lines is the {\em  incompatibility\/}
 or {\em defect density\/}
\begin{equation}
  \eta_{ij} ({\bf x}) =  \epsilon _{ikl}  \epsilon _{jmn} \partial _k
      \partial _m u_{ln}^P({\bf x}).
\label{@inc}\end{equation}
It can be decomposed into disclination and dislocation density as
follows  \cite{GFCM}:
\begin{equation}
\eta _{ij}({\bf x}) \!= \! \theta _{ij}({\bf x})\!+ \!
\frac{1}{2} \partial _m \left[
      \epsilon _{min}  \alpha _{jn}({\bf x})\! +\! (i\!\leftrightarrow\! j) \!- \! \epsilon _{ijn}
 \alpha _{mn}({\bf x})\right].
\label{@etaten}\end{equation}

 This tensor is symmetric and conserved
\begin{equation}
 \partial _i \eta_{ij}({\bf x}) = 0,
\label{@}\end{equation}
again a Bianchi identity of the defect
 system.

It is useful to separate from the
dislocation density
(\ref{DLD})
the contribution from the disclinations
which causes the nonzero right-hand side of
(\ref{DED}).
Thus we
 define
a {\em pure
dislocation density\/}
\begin{equation}
 \alpha^{b}_{ij}({\bf x})\equiv
 \alpha_{ij}({\bf x})-
 \alpha^{\Omega}_{ij}({\bf x})
\label{@}\end{equation}
which satisfies
 $\partial _i\alpha^{b}  _{ij}=0$.
Accordingly, we split
\begin{equation}
 \eta _{ij}({\bf x})=
 \eta _{ij}^{b}({\bf x})+
 \eta _{ij}^{\Omega}({\bf x}),
\label{@}\end{equation}
where
\begin{equation}  \!\phantom{x}
 \eta _{ij}^{b}({\bf x})\!=\!
\frac{1}{2}\left[
      \epsilon _{min}  \alpha _{jn}^{b}({\bf x}) + (i\leftrightarrow j) -  \epsilon _{ijn}
 \alpha _{mn}^{b}({\bf x})\right]\!, ~\!\!\!\!\!\!
\label{@}\end{equation}
and the pure disclination part of the defect tensor
looks like
(\ref{@etaten}), but with superscripts
$ \Omega $ on
$ \eta _{ij}$ and $ \alpha _{ij}$.

The tensors $ \alpha _{ij},\,  \theta _{ij }$, and $\eta_{ij}$ are
linearized versions of important geometric tensors in the {\em
Riemann-Cartan space\/} of defects, a noneuclidean space
  with
curvature and torsion.
Such a space can be generated from a flat space
by a plastic distortion,
which is mathematically
represented by a
{\em nonholonomic\/} mapping  \cite{NHOL,NHOL2}
%
$  x_i \rightarrow  x_i + u_i({\bf x}).$
%
Such a mapping is nonintegrable.
The displacement fields and their first derivatives
fail to satisfy the Schwarz integrability criterion:
\begin{eqnarray}  \!\!\!\!\!\!
\left(\partial _i \partial _j \!-\! \partial _j \partial _i\right)
     u({\bf x})   \neq  0 ,~~
\left( \partial _i \partial _j \!-\! \partial _j \partial _i\right)
    \partial _k u_l ({\bf x})  \neq  0.
\label{@}\end{eqnarray}
The
metric
and
the affine connection
of the geometry in the plastically distorted space are
$g_{ij} = \delta _{ij}+ \partial _i u_j + \partial _j u_i
$
and
%
$ \Gamma _{ijl} = \partial _i \partial _j u_l,$
%
respectively.
The noncommutativity of the derivatives in front
 of $u_l({\bf x})$ implies a nonzero torsion,
the torsion tensor being
%
$S_{ijk}\equiv ( \Gamma _{ijk}- \Gamma _{jik})/2.$
%
The dislocation density
$ \alpha _{ij}$ is
equal to
%
$  \alpha _{ij}=\epsilon _{ikl} S_{klj},$
%

The noncommutativity of the derivatives in front
 of $\partial _k u_l({\bf x})$ implies a nonzero
curvature.
The disclination density $ \theta _{ij}$ is the Einstein
 tensor
%
$\theta _{ij}=
R_{ji} -\frac{1}{2} g_{ji}R
$ 
of this Einstein-Cartan
defect geometry.
The tensor $\eta_{ij}$, finally, is the
Belinfante symmetric  energy momentum tensor,
 which is defined in terms of the canonical
 energy-momentum
tensor  and the spin density by a relation just like
(\ref{@etaten}).
    For more details on the
geometric aspects
see
 Part IV in Vol. II of
 \cite{GFCM},
where the full one-to-one
correspondence between
defect systems and Riemann-Cartan geometry is developed as well as
a
gravitational
theory
 based on this analogy.

Let us now show how linearized gravity emerges
from the energy
(\ref{@elen}).
For this we
eliminate the jumping
surfaces
in the defect gauge fields
from the partition function by introducing
 conjugate variables and
associated stress gauge fields. This is done by
rewriting the elastic action of defect lines as
\begin{eqnarray}
 \!\!\!\!\!\!\!\!\! E \!=\! \int d^3 x \left[\! \frac{1}{4\mu} \!
\left(\sigma _{ij}^2\!-\!\frac{ \nu }{1+ \nu }
\sigma _{ii}^2\right)
\!+\! i \sigma _{ij}
(u_{ij}
\!-\!u_{ij}^P)
\right]
,
\label{@EN0}\end{eqnarray}
where $ \nu \equiv  \lambda /2( \lambda +\mu)$
is Poisson's ratio,
and forming the partition function,
integrating
the Boltzmann factor $e^{-E/k_BT}$
 over $\sigma_{ij},  u_i$,
and summing over all jumping surfaces $S$
in the plastic fields.
The integrals
over  $u_i$ yield
the conservation law
%
$   \partial _i  \sigma  _{ij} = 0.$
%
This can
 be enforced as a Bianchi identity
 by
 introducing a stress gauge field
$h_{ij}$ and writing
%
$  \sigma _{ij} = G_{ij}\equiv
\epsilon _{ikl}\epsilon _{jmn}
\partial _k
\partial _m h_{ln}.$
%
The double curl on the right-hand side
 is
recognized as
 the Einstein tensor
in the geometric description of
stresses, expressed in terms of
a small deviation n $h_{ij}\equiv
g_{ij}- \delta _{ij}$
of the metric
from the
flat-space form.
Inserting $G_{ij}$
into (\ref{@EN0})
and using
(\ref{@inc}),
we can
replace the energy in the partition function
by
$
 E =E^{\rm stress}
 +E^{\rm def}$ where
\begin{equation}
E^{\rm stress}
 \!+\!E^{\rm def}
\! \equiv \!
 \int d^3 x \left[ \frac{1}{4\mu}
\left(G _{ij}^2\!-\!\frac{ \nu }{1+ \nu }
G _{ii}^2\right)
 +i
h _{ij} \eta _{ij}
\right],
\label{@ENN}\end{equation}
where the defect tensor
 (\ref{@etaten})
has the decomposition
\begin{equation}\!\!\!\!\!\!
 \eta _{ij}
=\eta^ \Omega _{ij}+
\partial _m
      \epsilon _{min}  \alpha ^b_{jn}
.
\label{@En3}\end{equation}
The defects have also  core energies which
has been ignored so far.
Adding these
for the dislocations
and ignoring, for a moment, the disclination part of the defect density
in (\ref{@En3}),
 we obtain
\begin{equation}
 E^{\rm disl}\!=\!i \int d^3 x \left(
      \epsilon _{imn}\partial _m
h _{ij}
 \alpha ^b_{jn}+\frac{ \epsilon _c}2{\alpha _{jn}^{b\,2}}
\right)
.      ~ \!\!\!\!
\label{@EN3}\end{equation}
We now assume
that the world crystal has undergone a transition
to a condensed phase in which dislocations are condensed.
To reach such a state, whose existence was
conjectured for two-dimensional crystals
in Ref.~ \cite{HN}, the model
requires a modification by an additional
rotational energy, as shown in  \cite{RST}
and verified by Monte Carlo simulations in
 \cite{RST2}.
The three-dimensional extension of the
model is described in  \cite{GFCM}, and analyzed in great detail
for the non-relativistic 2+1D quantum-fluid in  \cite{Zaanen}.

The condensed phase is described by a
partition function in which
the discrete sum over the pure dislocation densities
in $\alpha ^b_{jn}$ is
approximated by
an ordinary functional integral.
This has been shown
in Ref.~\cite{NHOL}.
The rule for summing
over closed-line ensembles
${l}_i({\bf x})= {\delta}_i({\bf x};L)$ in a
 phase
where the lines have proliferated is \cite{NHOL,NHOL2}
\begin{equation}
\int d^3l\, \delta ({\partial}_i { l}_i)\,e^{- \epsilon _c
 {l}_i^2/2+i  {l}_i
{a}_i}=e^{-{a}_{Ti}^{\,\,\,2}/2  \epsilon _c },
\label{@intrule}
\end{equation}
where
$a_{Ti}\equiv   -
i \epsilon _{ijk}\partial _ja_k/ \sqrt{-{\BF \partial}^2}$.
The Boltzmann
factor  resulting in this way from
                        $E^{\rm stress}$ plus
(\ref{@EN3})
has now
the energy
\begin{equation}
 E' = \int d^3 x \left[ \frac{1}{4\mu}
\left(G _{ij}^2\!-\!\frac{ \nu }{1+ \nu }
G _{ii}^2\right)
 +
\frac{1}{2 \epsilon _c}
G _{ij}\frac{1}{-{\BF \partial}^2}
G _{ij}
\right]
.
\label{@EN31}\end{equation}
The second term implies  a Meissner-like screening
of the initially confining gravitational forces between
the disclination part of the defect tensor
to Newton-like forces.
For distances longer than the Planck
scale,
we may
ignore the stress term
and find the effective gravitational action
for the disclination part of the defect tensor:
\begin{equation}
 E \approx
\int d^3 x \left(
\frac{1}{2 \epsilon _c}
G _{ij}\frac{1}{-{\BF \partial}^2}
G _{ij}+i h_{ij}\eta^ \Omega _{ij}
\right)
.
\label{@EN32}\end{equation}
 A path integral
over $h_{ij}$ and a sum over all line ensembles applied to the
Boltzmann factor $e^{-E/\hbar}$ is a simple Euclidean model of
pure quantum gravity. The line fluctuations of $\eta^ \Omega
_{ij}$ describe a fluctuating Riemann geometry perforated by a
grand-canonical ensemble arbitrarily shaped lines of curvature. As
long as the loops are small they merely renormalize the first term
in the energy (\ref{@EN32}). Such effects were calculated in
closely related theories in great detail in Ref.~ \cite{KM}. They
also give rise to post-Newtonian terms in the above linearized
description of the Riemann space.

We may now add matter to this gravitational environment \cite{padua}.
It is coupled by the usual Einstein interaction
\begin{equation}
 E^{\rm int} \approx
\int d^3 x \, h_{ij} T^{ij} , \label{@EN4}\end{equation} where  $
T ^{ij} $ is the symmetric Belinfante energy momentum tensor of
matter. Inserting for $G_{ij}$ the double-curl of $h_{ij}$ we see
that the energy (\ref{@EN32}) produces the correct Newton law if
the core energy is $ \epsilon _c=8\pi G$, where $G$ is Newton's
constant.

Note that the condensation process of dislocations has led to a
pure Riemann space without torsion. Just as a molten crystal shows
residues of the original crystal structure only at molecular
distances, remnants of the initial
 torsion could be observed only near the Planck scale.
This explains
why present-day
 general relativity
requires only a Riemann space, not a Riemann-Cartan space.

 In the non-relativistic context, a
dislocation condensate is characteristic for a nematic liquid
crystal, whose order is translationally invariant, but breaks
rotational symmetry (see  \cite{HN,GFCM} in
two dimensions and
 \cite{Zaanen} in the 2+1-dimensional quantum theory). The Burgers
vector of a dislocation is a vectorial topological charge, and
nematic order may be viewed as an ordering of the Burgers vectors
in the dislocation condensate. Initially,
a nematic order would
break the low energy Lorentz-invariance of space-time. We may,
however, imagine the stiffness of the directional field of
Burgers vectors to be so low that, by the criterion of
Ref.~ \cite{KleiCrit}, they undergo a Heisenberg-type of
phase transition into a directionally disordered phase
 in an environment with only a few disclinations \cite{Zaanen}.

In three dimensions, dislocations and disclinations are
line-like. This has the pleasant consequence, that they can  be
described by the disorder field theories developed in  \cite{rem2}
in which the proliferation of dislocations follows the typical
Ginzburg-Landau pattern of the field expectation acquiring a
nonzero expectation value. A cubic interaction becomes isotropic
in the continuum limit  \cite{rem3} (this is the famous
fluctuation-induced symmetry restoration of the Heisenberg
fixed-point in a $\phi^4$-theory with O(3)-symmetric plus cubic
interactions  \cite{KS}). The disordered, isotropic phase is just a
physical realization  of what has
been called a {\em topological} form  of nematic order by Lammert
et al.  \cite{Lammert} in their $O(3)/Z_2$ gauge theory of
nematic order: disclinations are massive although rotational symmetry
is unbroken.

The above description of defects was formulated in what has been
named tangential approximation to the Euclidean group  \cite{GFCM},
in which the discrete rotations are treated as if they took place
in the tangential place with arbitrary real Franck vectors
$\Omega^i$ [see Eq.~(\ref{DLD})]. In a more accurate formulation,
the nonabelian nature of the rotations and the quantization of
$\Omega_i$ must be taken into account. Their discreteness is
certainly remembered in the nematic phase, even if the directions
of the Burgers vectors become disordered (see also the discussion
in Ref.~ \cite{Bais}). This implies that there are elements of
quantized curvature fluctuating in spacetime. Fortunately, this
does not introduce any observable consequence at  presently
accessible length scales since these
fluctuations mainly renormalize the basic curvature energy in
(\ref{@EN32}), as discussed before.

In one regard, our
 relativistic nematic fluid
departs radically from the nematics formed from
non-relativistic matter. In the energy (\ref{@EN31}),
all components of the stress tensor
$G_{ij}$ have acquired the same mass which makes both shear- and
compression stresses short-ranged. This is in contrast to
non-relativistic crystalline matter, where shear rigidity is
associated with translational symmetry breaking, and the
dislocation condensate gives a Meissner-Higgs mass only to shear
modes. Accordingly, the non-relativistic quantum
nematic supports a massless compressional mode which
is just the phase mode of the superfluid \cite{Zaanen}. The absence
of compressional rigidity in the relativistic nematic is in fact
quite natural. In order to decouple compression
from the dislocation condensate an independent
dynamical constraint is required: the glide condition, implying
that dislocations only propagate in the directions of their
(spatial) Burgers vector. Orthogonal directions would involve an
excessive quantum migration of atoms in the crystal. Glide can
only meaningfully be defined in a non-relativistic spacetime,
with the consequence that compressional rigidity has to disappear
in the relativistic fluid.

In conclusion, we have demonstrated that the spacetime of
general relativity can in principle be interpreted in terms
of a material medium similar to
 a relativistic
quantum-nematic fluid. Our construction is explicit in three
Euclidean space-time dimensions. The generalization to four
dimensions changes mainly the geometry of the defects.
These become world sheets, and a second-quantized
disorder field description of surfaces has not yet been found. But
the approximation of representing a sum over dislocation surfaces
in the proliferated phase as an integral as in
Eq.~(\ref{@intrule}) does remain valid, so that the above line of
arguments will survive, this being a natural generalization of the
Meissner-Higgs mechanism.

One might want to view our finding as a
metaphor, offering an alternative perspective on the nature of
spacetime, which might be of use in the greater context of the
quantum gravity enigma. However, despite its simplicity, it
cannot be a-priori excluded that the idea has a more literal
meaning. It implies that Lorentz invariance is emergent and
this alone cures the singularities of quantum gravity. Next,
our model has automatically a vanishing cosmological constant.
Since the atoms in the crystal are in equilibrium, the pressure
is zero. This explanation is similar to that given by Volovik  \cite{Volovik1}
with his helium droplet  analogies. Finally, the theory predicts
that at the Planck  scale,
 the disclination sources of curvature should become visible. As an
attractive feature for string theorists, these will be world sheets.
However, the high energy  properties will be completely different
from the common strings, because these surfaces behave nonrelativistically
as the energies approach the Planck scale. Accordingly, they will neither
exhibit the recurrent particle spectra at arbitrary multiples of the Planck
mass, nor the deep conceptual problems associated with explaining the
low-energy universe \cite{Banks}. The deviations from relativity at
high energies or short distances associated with a literal
interpretation of the `world nematic' may come into
experimentalists reach in the not too distant future.

%


\end{document}